\documentclass[aps,twocolumn,showpacs,superscriptaddress]{revtex4}% Physical Review
\usepackage{graphicx}% Include figure files
\usepackage{bbm}
\usepackage{amsmath}% More mathematical features
\usepackage{amssymb}     
\usepackage{epsfig}
\usepackage{pstricks,pst-grad}% PsTricks

\begin{document}

%\preprint{ECM}

%\title{Anomaly in the metastable states in critical hysteresis}

\title{Numerical study of metastable states in the $T=0$ RFIM}

\author{F.J.~P\'erez-Reche}
%
%\affiliation{Laboratoire de M\'ecanique des Solides, Ecole Polytechnique, 91128 Palaiseau, France}
\affiliation{Laboratoire de Physique Th\'eorique de la Mati\`ere Condens\'ee, Universit\'e Pierre et Marie Curie\\ 4 place Jussieu, 75252 Paris Cedex 05, France}
\affiliation{ Laboratoire de M\'ecanique des Solides, Ecole Polytechnique, 91128 Palaiseau, France}

\email{perez@lms.polytechnique.fr}
\affiliation{Dipartamento di Metodi e Modelli Matematici per le Scienze Applicate, Universit\`a di Padova \\ Via Trieste 63, 35121 Padova, Italy}

\author{M.L.~Rosinberg}
\affiliation{Laboratoire de Physique Th\'eorique de la Mati\`ere Condens\'ee, Universit\'e Pierre et Marie Curie\\ 4 place Jussieu, 75252 Paris Cedex 05, France}
%
%\email{mlr@lptmc.jussieu.fr}

\author{G.~Tarjus}
\affiliation{Laboratoire de Physique Th\'eorique de la Mati\`ere Condens\'ee, Universit\'e Pierre et Marie Curie\\ 4 place Jussieu, 75252 Paris Cedex 05, France}
%
%\email{tarjus@lptmc.jussieu.fr}

\date{\today}

\begin{abstract}
  We study numerically the number of single-spin-flip stable states in the $T=0$ Random Field Ising
  Model (RFIM) on random regular graphs of connectivity $z=2$ and $z=4$ and on the cubic lattice.
  The annealed and quenched complexities (i.e. the entropy densities) of the metastable states with
  given magnetization are calculated as a function of the external magnetic field. The results show
  that the appearance of a (disorder-induced) out-of-equilibrium phase transition in the
  magnetization hysteresis loop at low disorder can be ascribed to a change in the distribution of the
  metastable states in the field-magnetization plane.
\end{abstract}

%64.60.My: Metastable phases

\maketitle

\section{Introduction}

\begin{figure}
\epsfig{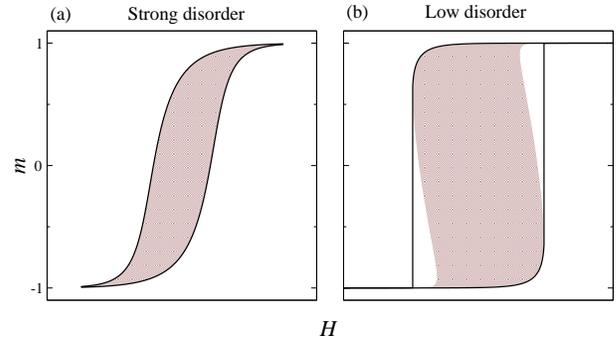}
  \caption{\label{Figintro} Putative distribution of the metastable states in the
    field-magnetization plane for the $T=0$ RFIM on the cubic lattice (or the Bethe lattice with $z=4$): (a) strong disorder regime (b) low disorder regime. The number of single-spin-flip
    stable states scales exponentially with the system size in the shaded region and is zero
    outside. The solid curve is the associated magnetization hysteresis loop. (Color on line.)}
\end{figure}

At low temperature, disordered systems may be trapped in long-lived
metastable states separated by large energy barriers that make
relaxation to thermal equilibrium unlikely on experimental time
scales. As a consequence, these systems, when driven by an external
force, display a jerky and hysteretic behavior which is associated to
irreversible jumps between the metastable states, events that are
called {\it avalanches}. The out-of-equilibrium behavior is thus
dictated by the properties of the local minima in the energy (or
free-energy) landscape and it would be useful to know {\it a priori}
their number and energy, or, in the case of magnetic systems, their
magnetization. However, such a computation is a nontrivial task, even
when the metastable states can be clearly identified, as it is the
case at zero temperature or in mean-field models. Consider for
instance the ferromagnetic Random Field Ising Model (RFIM) at $T=0$
which is a simple prototype of a spin system with avalanche behavior
and has been extensively studied in recent years
\cite{Sethna_review2004}. In this case, one would like to relate the
hysteretic response of the system to an externally varying magnetic
field to the distribution of the metastable states in the
field-magnetization plane. In general, one expects the typical number
of metastable states to scale exponentially with the system size
inside the saturation hysteresis loop and to vanish
outside\cite{Detcheverry2005}. At low disorder, however, the existence
of an out-of-equilibrium phase transition, characterized by a
macroscopic jump in the hysteresis loop \cite{Sethna1993}, strongly
suggests that the typical number of states also vanishes in some
region {\it inside} the loop, as depicted schematically in Fig.
\ref{Figintro}(b). This feature may also explain the behavior observed
when the system is driven by the magnetization instead of the magnetic
field\cite{Illa2006a,Illa2006b}.

In a recent paper\cite{Detcheverry2005}, this issue was investigated analytically in the case of the RFIM on random regular graphs of fixed connectivity $z$ (akin to the Bethe lattice~\cite{Mezard2001}) for which the out-of-equilibrium phase transition occurs when $z\ge 4$ \cite{Dhar1997}. The annealed complexity $\Sigma_A(m;H)$ and the quenched complexity $\Sigma_Q(m;H)$ of the single-spin-flip stable states (i.e., the entropy densities associated to the average and typical number of states, $\overline{{\cal N}(m;H)}$ and $\exp[\overline{\ln {\cal
N}(m;H)}]$, respectively) were studied as a function of the magnetization (per spin) $m$ and the external field $H$. The quenched complexity $\Sigma_Q(m;H)$, however, could only be obtained analytically for $z=2$ (i.e., in one dimension) and therefore the relationship between the phase transition in the hysteresis loop and the distribution of the typical states could not be investigated. On the other hand, $\Sigma_A(m;H)$ was computed for several values of $z$ but it was shown that its behavior does not reflect the typical behavior of the system; the difference may even be qualitative as for the change in the concavity of $\Sigma_A(m;H)$ which already occurs for $z=3$.

In the present work, we revisit the problem by performing numerical computations of $\Sigma_A(m,H) $ and
$\Sigma_Q(m,H)$ on both random graphs and the cubic lattice. Since exact enumeration is only possible for
systems of very small size, we also use more sophisticated statistical methods that allow us to
consider larger systems and to make sensible extrapolations of the data to the thermodynamic limit despite the fact that the number of metastable states grows exponentially. The case $z=2$
for which complete analytical results are available is used as a benchmark.

The rest of the paper is organized as follows. In section \ref{Sec.Model}, we introduce the model,
define the quantities that we want to compute, and briefly recall the main results of
Ref.~\onlinecite{Detcheverry2005}. In section~\ref{sec:Complex-Finite}, we explain how to properly define
the complexities in finite systems, and, in section \ref{sec-Numerical}, we describe the numerical
methods. The results for the $z=2$ random graphs are presented in section \ref{z2} and those for
$z=4$ in section \ref{z4}. The case of the cubic lattice, for which no analytical
treatment is possible, is studied in section \ref{Cubic-Lattices}. Finally, in section
\ref{Conclusion}, we give some general remarks and conclude.

\section{Model}
\label{Sec.Model}
The RFIM is defined as a set of $N$ Ising spins $\{s_i=\pm 1,
i=1,\dots,N\}$ placed on the vertices of a graph, with Hamiltonian
\begin{equation}
\label{Eq-Model-1} {\cal H}=-\frac{J}{2}\sum_{i,j=1}^N
n_{ij} s_i s_j-H\sum_{i=1}^N s_i-\sum_{i=1}h_i s_i.
\end{equation}
$J$ is a ferromagnetic coupling constant (we take $J=1$ in the simulations) and $n_{ij}$
is a connectivity matrix that defines the topology
of the graph ($n_{ij}=1$ if the vertices $i$ and $j$ are
connected and $0$ otherwise). In the following we shall consider either random regular graphs with fixed connectivity $z=2$ and $z=4$ or the $3d$ cubic lattice. $H$ is a uniform external field and the quenched local fields $h_i$ are independent random variables drawn from a Gaussian probability distribution with zero mean and standard deviation $\Delta$ which parametrizes the amount of disorder in the system.

Hysteresis and avalanches in the RFIM have been extensively studied in the past years using the standard zero-temperature Glauber dynamics\cite{Sethna1993,Perkovic1999,PerezReche2003,Colaiori2004}. A state is then (meta)stable when its energy cannot be decreased by flipping a single spin.
Although more general dynamics may be also considered, in which a configuration is stable if its energy cannot be decreased by flipping any subset of $1,2,...,k$ spins\cite{Newman-Stein1999}, we shall only focus on 
the 1-spin-flip stable states that we simply call metastable (all other states are thus unstable with respect to the dynamics)

From Eq.~(\ref{Eq-Model-1}), the change in the energy due to the
flip of a single spin $s_i$ is
\begin{equation}
\label{Eq-Model-3} 
\Delta {\cal H}(s_i \rightarrow -s_i) = 2s_i f_i,
\end{equation}
where $f_i=J\sum_{j \neq i} n_{ij} s_j+h_i+H$ is the effective local
field acting on $s_i$. The sum extends over the $z$ spins connected
with $s_i$ (because the random field distribution is continuous, 
the probability of having $f_i=0$ is zero).  According to Eq. (\ref{Eq-Model-3}), all the spins $s_i$ in a
metastable configuration are aligned with their local field, i.e.,
\begin{equation}
\label{Eq-Model-5} s_i f_i >  0 .
\end{equation}
As will be discussed below, it is also useful to introduce a
``misalignment'' parameter $\mu$ ($0\leq \mu\leq 1$) defined as
\begin{equation}
\label{Eq-Model-6}
\mu=\frac{1}{N}\sum_{i=1}^{N} \Theta(-s_i f_i),
\end{equation}
where $\Theta(x)$ is the Heaviside step function ($\Theta(x)=1$ if $x>0$ and $0$ if $x<0$). $\mu$ is the fraction of spins in a given configuration that are not aligned
with their local field (metastable states thus correspond to $\mu=0$).

In the following, the two quantities that describe macroscopically a disorder realization are 
$\hat{\cal N}(m,\mu;H)$, the number of states at field $H$ with magnetization $m=(1/N)\sum_i s_i$ and misalignment
$\mu$, and ${\cal N}(m;H)=\hat{\cal N}(m,0;H)$, the number of metastable states with magnetization
$m$. These quantities are averaged over a large number of samples
characterized by a different set of random fields and random graphs.  Since both
$\overline{{\cal N}(m;H)}$, the average number of metastable states, and $\exp[\overline{\ln {\cal
    N}(m;H)}]$, the typical number, are expected to scale exponentially with $N$ in some region of
the $H-m$ plane, the interesting quantities are the corresponding annealed and quenched
complexities
\begin{align}
\label{Intro.Eq.1.1}
\Sigma^{\infty}_A(m;H)=\lim_{N \rightarrow \infty} \frac{1}{N}\ln \overline{ {\cal N}(m;H)},\\ 
\label{Intro.Eq.1.2}
\Sigma^{\infty}_Q(m;H)=\lim_{N \rightarrow \infty} \frac{1}{N} \overline{ \ln {\cal N}(m;H)},
\end{align}
where the superscript `$\infty$' indicates that these quantities refer
to the thermodynamic limit. As shown in Ref.~\onlinecite{Detcheverry2005},
the probability of finding metastable states outside the average
hysteresis loop vanishes when $N\rightarrow \infty$. Accordingly,
$\Sigma^{\infty}_Q(m;H)=-\infty$ (or is not defined) in this region
whereas $\Sigma^{\infty}_A(m;H)$ may be positive because of the
existence of nontypical metastable states.  In other words, the
contour $\Sigma^{\infty}_A(m;H)=0$ overestimates the size of the
actual hysteresis loop. The situation {\it inside} the loop is more
complicated, depending on the connectivity $z$ of the random graphs
(or the spatial dimension) and the disorder strength.  For $z=2$ (or
in one dimension), $\Sigma^{\infty}_A$ and $\Sigma^{\infty}_Q$ are
concave functions of $H$ and $m$ for all values of $\Delta$. On the
other hand, for $z \geq 3$ and $\Delta$ small enough,
$\Sigma^{\infty}_A$ becomes a nonmonotonic function of $m$ in some
range of the field (Fig.  6 in Ref.~\onlinecite{Detcheverry2005}) and
$m_A^{\infty}(H)$, the (annealed) average magnetization of the states
(corresponding to the maxima of $\Sigma^{\infty}_A(m;H)$), may display
a discontinuity (Fig.  9 in Ref.~\onlinecite{Detcheverry2005}). It is likely
that the same is true on euclidean lattices in $3$ and higher
dimensions. The corresponding behavior of $\Sigma^{\infty}_Q(m;H)$ is
still unknown but, as discussed in Ref.~\onlinecite{Detcheverry2005}, the
existence of a jump in the hysteresis loop at the coercive field (for
$z\geq 4$ or $d\geq 3$ at low disorder) suggests that the curve
$\Sigma^{\infty}_Q(m;H)=0$ has a reentrant part, as depicted in Fig.
\ref{Figintro}(b). This implies that $\Sigma^{\infty}_Q(m;H)$ has at
least two maxima as a function of $m$ in a certain range of the field.
It is also possible that the typical magnetization of the metastable
states (corresponding to the maximum of $\Sigma^{\infty}_Q(m;H)$) has
a discontinuity at a certain value of $H$.  This is the general
scenario that we try to confirm numerically in the present work.

\section{Definition of the complexities in finite-size systems}
\label{sec:Complex-Finite}

Since numerical simulations are performed in finite systems, one must define the corresponding
$N$-dependent complexities in such a way that one can properly extrapolate the results to the
thermodynamic limit.

It is quite natural to define the annealed complexity in finite systems as
\begin{equation}
\label{Complex.Eq.1}
\Sigma_A(m;H,N)=\frac{1}{N} \ln \overline{{\cal N}(m;H,N)},
\end{equation}
so that the average number of metastable states is just $\overline{{\cal N}(m;H,N)}=\exp[N\Sigma_A(m;H,N)]$. A similar definition for the quenched complexity would be $\Sigma_Q(m;H,N)=(1/N) \overline{\ln {\cal
    N}(m;H,N)}$. This is not convenient, however, because the average of the logarithm diverges
as soon as ${\cal N}(m;H,N)$ is zero in some sample, a situation that always occurs in a system of small size. One could imagine to circumvent the problem by arbitrarily setting
$\ln({\cal N}(m;H,N))=0$ in this case (see, e.g., Ref.~\onlinecite{Cherrier-Dean-Lefevre2003}). This
should be unimportant in the thermodynamic limit since it concerns a number of states that
is nonexponential in system size. However, we recall that we are especially interested in the region
where the quenched complexity is very small and vanishes when $N\rightarrow \infty$. This procedure
would thus introduce an unacceptable bias.  The alternative solution that we propose is
inspired by the analytical calculation of Ref.~\onlinecite{Detcheverry2005}, introducing the function $\Lambda_{\alpha}(g;H,N)$ defined by the following  Legendre-Fenchel transform
\begin{equation}
\label{Complex.Eq.2}
\Lambda_{\alpha}(g;H,N)=\max_m\left\{\frac{1}{N} \ln {\cal N}_{\alpha}(m;H,N)+gm \right\},
\end{equation} 
where $g\in \mathbbmss{R}$  (we provisionally add the subscript $\alpha$ to recall that ${\cal N}(m;H,N)$ is
sample-dependent). Note that we take the maximum of the expression inside the parenthesis with
respect to $m$, which cures the problem of ${\cal N}_{\alpha}(m;H,N)$ being zero for some values of
$m$.

\begin{figure}
\epsfig{file=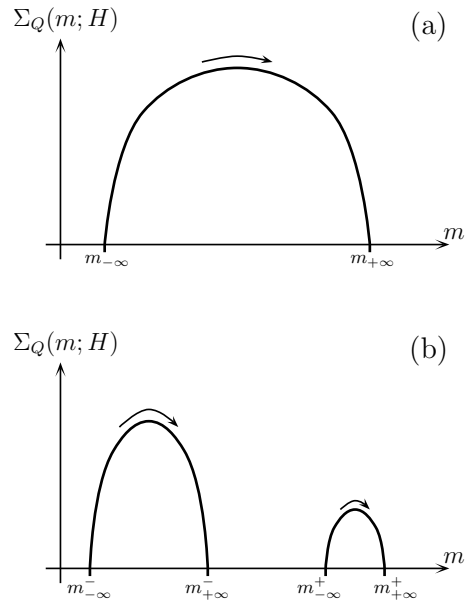, width=6cm,clip=}
  \caption{\label{Fig1_5} Schematic behavior of the quenched complexity in the presence of a disorder-induced phase transition (e.g., on random graphs with $z=4$). (a) $\Sigma_Q(m;H)$ has a single maximum and is positive in the range $m^{}_{-\infty}(H)\leq m \leq m^{}_{+\infty}(H)$ (b) $\Sigma_Q(m;H)$ has two maxima
and is positive in the range  $m^{-}_{-\infty}(H)\leq m \leq m^{-}_{+\infty}(H)$ and $m^{+}_{-\infty}(H)\leq m \leq m^{+}_{+\infty}(H)$. $m$ and $\Sigma_Q(m;H)$ are parametrized by the Legendre variable $g$ which increases from $-\infty$ to $+\infty$ as indicated by the arrows.}
\end{figure}
Let us first consider the simplest situation where ${\cal N}_{\alpha}(m;H,N)$ has a single maximum (i.e., is concave). Then, for a given $g$, the function $(1/N) \ln {\cal
  N}_{\alpha}(m;H,N)+gm$  (or, equivalently, $e^{Ngm} {\cal N}_{\alpha}(m;H,N)$) has a single maximum at some magnetization $m_{\alpha}^{}(g;H,N)$ and
\begin{equation}
\label{Complex.Eq.3}
\Lambda_{\alpha}(g;H,N)=\frac{1}{N} \ln {\cal N}_{\alpha} (m_{\alpha}^{};H,N)+g m_{\alpha}^{}.
\end{equation}
The average over disorder yields
\begin{equation}
\label{Complex.Eq.3b}
\Lambda(g;H,N)=\frac{1}{N} \overline{\ln {\cal N}_{\alpha} (m_{\alpha}^{};H,N)}+g m 
\end{equation}
where $m(g;H,N)\equiv \overline{m_{\alpha}^{}(g;H,N)}$. This leads to define the size-dependent
quenched complexity as
\begin{equation}
\label{Complex.Eq.4}
\Sigma_Q(m;H,N)=\Lambda(g;H,N)-gm, 
\end{equation}
neglecting terms of order $1/N$ (as one expects
that $m_{\alpha}^{}$ deviates from it average value by terms of order $1/\sqrt N$) and considering both $m$ and $\Sigma_Q(m;H,N)$ as functions of the independent parameter $g$ through the
relation $m=m(g;H,N)$. Clearly, one has
$\Sigma_Q(m;H,N)\rightarrow \Sigma^{\infty}_Q(m;H)$ when $N\rightarrow \infty$, and $\Lambda$ and $\Sigma_Q$ are related by the standard Legendre relations $\partial \Lambda/\partial g=m$ and $\partial \Sigma_Q/\partial m=-g$ (neglecting again terms of order $1/N$ to derive the second relation). Varying $g$ from $-\infty$ to $+\infty$ then traces out the curve
$\Sigma_Q(m;H,N)$ {\it vs.} $m$, as depicted schematically in Fig.~\ref{Fig1_5}(a). Moreover, one readily sees from Eq. (\ref{Complex.Eq.2}) that
the limit $g\rightarrow -\infty$ (resp. $g\rightarrow +\infty$) selects in each sample $\alpha$ the smallest (resp. largest) magnetization for which ${\cal N}_{\alpha} (m;H,N)$ is strictly positive.  As shown in Ref.~\onlinecite{Detcheverry2005}, these two values of $m$ correspond to the lower and upper branch of the
hysteresis loop, respectively.  Therefore, the two curves obtained by plotting $m_{\pm
  \infty}^{}(H,N)=\lim_{g\rightarrow \pm \infty}m(g;H,N)$ as a function of $H$ define
the two branches of the average hysteresis loop. On the other hand, the typical (i.e. the most likely) value of the magnetization of the metastables states is described by the curve $m(g=0;H,N)$ which corresponds to the locus of the maximum of $\Sigma_Q$.

Now, consider the situation where ${\cal N}_{\alpha}(m;H,N)$ has two
maxima associated to two separated lobes (see e.g. Fig.~\ref{Fig9_5}
in section \ref{Sigma_Q-z4}). As discussed previously, such a
situation is likely to occur in a certain range of $H$ when there
exists an out-of-equilibrium phase transition, for instance on the
$z=4$ Bethe lattice at low disorder.  One expects the quenched
complexity to display the same behavior, as depicted in Fig.~2(b), but
this information cannot be extracted from Eqs.
(\ref{Complex.Eq.2}-\ref{Complex.Eq.4}). In this case, it is
convenient to treat the two maxima separately by introducing two
Legendre-Fenchel transforms:
\begin{align}
\label{Complex.Eq.5}
\Lambda_{\alpha}^{-}(g;H,N)&=\max_{m<m_{th}}\left\{\frac{1}{N} \ln {\cal N}_{\alpha}(m;H,N)+gm \right\}\nonumber\\
\Lambda_{\alpha}^{+}(g;H,N)&=\max_{m>m_{th}}\left\{\frac{1}{N} \ln {\cal N}_{\alpha}(m;H,N)+gm\right\},
\end{align}
where $m_{th}$ is a certain threshold in magnetization which does not depend on the disorder realization
and which is chosen so to lie between the two maxima for as many realizations as
possible. In consequence, the size-dependent quenched complexity also splits into two parts,
\begin{align}
\label{Complex.Eq.6}
%\Sigma_Q^{\gtrless}\left(m^{\gtrless};N,H\right)=\Lambda^{\gtrless}-gm^{\gtrless}.
\Sigma_Q^{-}\left(m;H,N)\right)&=\Lambda^{-}(g;H,N)-gm^{-}\nonumber\\
\Sigma_Q^{+}\left(m;H,N)\right)&=\Lambda^{+}(g;H,N)-gm^{+}
\end{align}
where $m^{-}(g;H,N)\equiv\overline{m_{\alpha}^{-}(g;H,N)}$ and $m^{+}(g;H,N)\equiv\overline{m_{\alpha}^{+}(g;H,N)}$. It is again useful 
to consider the magnetization and the complexity $\Sigma_Q^{-}$ or $\Sigma_Q^{+}$ as functions of the independent parameter $g$ through the
relations $m=m^{-}(g;H,N)$ or $m=m^{+}(g;H,N)$. Of course, one recovers the preceding formulas with only one function $\Lambda$ or $\Sigma_Q$ if $\mid m_{th}\mid>1$.

It is also interesting to consider the limits $g\rightarrow \pm
\infty$. Now $m^{-}_{-\infty}(H,N)=\lim_{g \rightarrow -\infty}
m^{-}(g;H,N)$ and $m^{+}_{+\infty}(H,N)=\lim_{g \rightarrow +\infty}
m^{+}(g;H,N)$ correspond to the lower and upper branches of the
hysteresis loop, respectively. On the other hand, in
Eqs.~(\ref{Complex.Eq.5}), the limit $g\rightarrow +\infty$ (resp.
$g\rightarrow -\infty$) selects in each sample the largest (resp.
smallest) magnetization in the interval $[-1,m_{th}]$ (resp.
$[m_{th},1]$) for which ${\cal N}_{\alpha}(m;H,N)>0$.  Therefore, if
$m^{-}_{\alpha}(g\rightarrow +\infty;H,N)< m_{th}<
m^{+}_{\alpha}(g\rightarrow -\infty;H,N)$, there is a gap in magnetization for which ${\cal N}_{\alpha}(m;H,N)=0$.  On the contrary,
if $m^{-}_{\alpha}(g\rightarrow +\infty;H,N)=
m^{+}_{\alpha}(g\rightarrow -\infty;H,N)$, the gap does not exist. In
finite systems, the width of the gap fluctuates from one sample to
another and the appropriate quantity to be analyzed is the average gap.

\section{Numerical methods }
\label{sec-Numerical}
We now detail the numerical methods that we have used to compute the number of metastable states
as a function of the magnetization and the external field.

\subsection{Exact enumeration}
\label{sec-Exact}
The simplest method for counting the number of metastable states in a given
disorder realization consists in generating
all the spin configurations  and selecting those which are metastable in a certain range of the field.
To do this, we determine the interval of stability $[H_{min},H_{max}]$
of each configuration $\mathbf{s}=\{s_i\}$. Eq.~(\ref{Eq-Model-5}) implies that
a spin $s_i=+1$ (resp. $s_i=-1$) is stable as long as $H \ge -J\sum_{j \neq i} n_{ij} s_j-h_i$
(resp. $H \le -J\sum_{j \neq i} n_{ij} s_j-h_i$). The fields $H_{min}$ and
$H_{max}$ correspond to the two situations of marginal stability for $\mathbf{s}$, i.e.,
\begin{align}
\label{Eq-Model-6.1}
H_{min} &= \max_{\{i|s_i=+1\}} \{-J\sum_{j \neq i} n_{ij} s_j-h_i\}\\
\label{Eq-Model-6.2} H_{max} &= \min_{\{i|s_i=-1\}} \{-J\sum_{j \neq i} n_{ij} s_j-h_i\} \ .
\end{align}
Accordingly, configurations with $H_{min} \leq H_{max}$ are stable when $H$ is in the range $[H_{min},H_{max}]$.  On the other hand, configurations with $H_{min} > H_{max}$ are always 
unstable since, whatever the value of $H$, at least the spin up corresponding to $H_{min}$ (or the spin down corresponding to $H_{max}$) is unstable.  Since the total number of
states grows like $2^N$, one of course cannot consider large systems. In the present work, the largest size investigated by this method is $N=26$ (with $5000$ disorder realizations). However, as will be seen below, some of the data can be reasonably extrapolated to the thermodynamic limit. Moreover, even in a small system, one can observe a clear connection between the shape of 
the hysteresis loop and the intervals of stability
$[H_{min},H_{max}]$ of the metastable states.

\begin{figure}
\epsfig{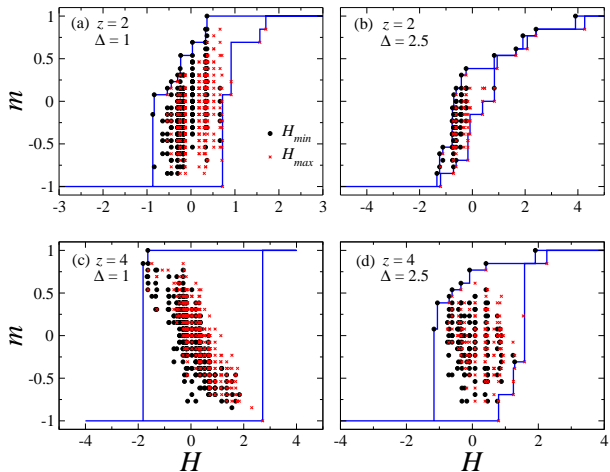}
\caption{\label{Fig1} Limits of stability $H_{min}$ (circles) and $H_{max}$ (crosses) of the
  metastable states in a single disorder realization on a random graph of size $N=26$: (a) $z=2$, $\Delta=1$, (b) $z=2$, $\Delta=2.5$ (c) $z=4$, $\Delta=1$ (d) $z=4$, $\Delta=2.5$.(Color on line.)}
\end{figure}

This is illustrated in Fig.~\ref{Fig1} where we plot all the $H_{min}$ and $H_{max}$ obtained in a single sample (namely, a realization of the random fields on a random graph of size $N=26$) for different connectivities $z$ and/or values of $\Delta$. For a given overall magnetization
$M=Nm$, there are in general several metastable states, each one characterized by a stability range $[H_{min},H_{max}]$ (for clarity, no link
between $H_{min}$ and $H_{max}$ has been drawn in the figure.) The
comparison of the distribution of these marginal fields in the $H$-$m$ plane
with the corresponding saturation hysteresis loop shows that:

(i) There are no metastable states outside the loop, as  expected theoretically\cite{Detcheverry2005}.

(ii) The hysteresis loop is a path in the $H$-$m$ plane that connects the extremal
  states, i.e. those which have the largest $H_{max}$ at a given $m$ 
  (on the ascending branch) or the smallest $H_{min}$ (on the descending branch). An avalanche (for
  instance on the ascending branch) occurs between the extremal state $\mathbf{s}$ at $m$ and the
  extremal state $\mathbf{s}'$ at $m'>m$ when all the stable states with intermediate magnetization
  $m<m''<m'$ have a $H_{max}$ which is smaller than $H_{max}(\mathbf{s})$. (In fact, this property is
  true in general for any pair of ordered states $\mathbf{s}$ and
  $\mathbf{s}'>\mathbf{s}$\cite{note1}). In other words,
  avalanches connect metastable states through a sequence of unstable (or, at most, marginally stable)
  intermediate configurations.  For $z=2$ (and both $\Delta=1$ and $\Delta=2.5$), one can see in
  Figs.~\ref{Fig1}(a) and (b) that there are indeed some intermediate marginal states along the
  magnetization discontinuities of the hysteresis loop. In contrast, for $z=4$, marginal
  intermediate states are only observed when $\Delta$ is large enough (Fig.~\ref{Fig1}(d)). At low disorder
  (Fig.~\ref{Fig1}(c)), there exists a significant region inside the loop where there are
  no metastable states at all. This suggests that the discontinuity in
  the hysteresis loop in the thermodynamic limit is related to the existence of a region inside the
  loop without any typical metastable states.

\subsection{Entropic sampling}
\label{sec-Entropic}

The second method that we have used to compute the number of metastable states is the entropic sampling Monte Carlo (MC) algorithm\cite{Lee1993,Shteto1997}, which is a variant of the multicanonical approach\cite{Berg1992}. This algorithm explores the configurational space with a biased sampling probability that favors the weakly degenerate macrostates and disfavor the highly degenerate ones. The final result is 
$\hat{\cal N}(m,\mu;H)$, with ${\cal N}(m,\mu=0;H)$ as a by-product. 

The practical implementation of the method is as follows.  Starting from an initial guess for $\hat{\cal N}(m,\mu;H)$ (for instance $\hat{\cal
  N}_0(m,\mu;H)=1$ for all $m$ and $\mu$), one improves the estimate iteratively via a sequence of Monte Carlo runs. The number of states $\hat{\cal N}_k(m,\mu;H)$  obtained after iteration (or ``stage'' \cite{Shteto1997}) $k$ is used to bias the acceptance probability in the next MC run (i.e.,  the transition $s
\rightarrow s'$ is accepted with the  probability $\min\{1,\hat{\cal N}_k(m,\mu;H)/\hat{\cal
  N}_k(m',\mu';H)\}$).  One then obtains a histogram
$\texttt{H}_{k+1}(m,\mu)$ of the frequency of the macrostates  and $\hat{\cal N}_k(m,\mu;H)$ is updated according to the rule
\begin{equation}
\label{Eq-Entropic-4}
\hat{\cal N}_{k+1}(m,\mu;H)=%
\begin{cases}
\hat{\cal N}_{k}(m,\mu;H)\;\; \text{if $\texttt{H}_{k+1}(m,\mu)=0$}\\
\hat{\cal N}_k(m,\mu;H) \cdot \texttt{H}_{k+1}(m,\mu) \;\; \text{otherwise}.
\end{cases}
\end{equation}

In principle, this method could yield the actual $\hat{\cal N}(m,\mu;H)$ after a single very long MC
run, even if the initial guess is very crude (choosing $\hat{\cal N}_0(m,\mu;H)=1$ implies that the states are first
sampled at random since all spin flips are accepted). The problem with this
brute-force procedure, however, is that the number of MC steps required to detect the macrostates that are
exponentially less degenerate than the other ones increases exponentially with $N$. In consequence,
it is better to update $\hat{\cal N}(m,\mu;H)$ iteratively so to disfavor the most degenerate macrostates
and facilitate the search of those which are less degenerate.  We have followed the strategy proposed
in Refs.~\cite{Lee1993,Shteto1997} which consists in starting with short MC stages and increasing
the length of the subsequent stages if the number of visited states decreases.  In this way,
the number of visited states increases and eventually saturates after a certain number of iterations $k^*$
that mainly depends on the system size and the length of the MC stages. Saturation indicates that all
the existing macrostates have been already visited (and the corresponding histogram is flat).  In this situation, the number $\hat{\cal
  N}_{k^*}(m,\mu;H)$ corresponding to non-visited states is set to $0$ (instead of the initial guess $1$) and one finally obtains
$\hat{\cal N}(m,\mu;H)$ from $\hat{\cal N}_{k^*}(m,\mu;H)$ via the rescaling 
\begin{equation}
\hat{\cal N}(m,\mu;H)= \frac{2^N}{\sum_{m,\mu} \hat{\cal  N}_{k^*}(m,\mu;H)} \hat{\cal  N}_{k^*}(m,\mu;H),
\end{equation}
which ensures that the sum $\sum_{m,\mu}\hat{\cal N}$ is equal to $2^N$.

\begin{figure}[h]
  \epsfig{file=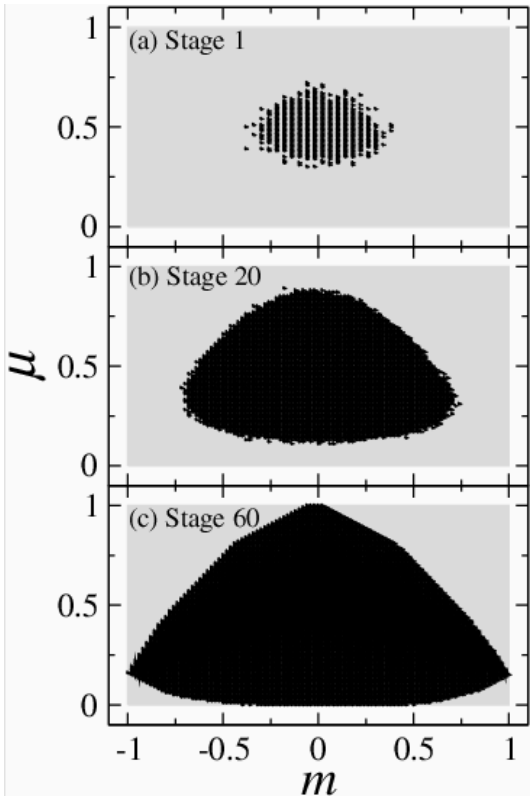, width=7cm,clip=}
  \caption{\label{Fig2} Entropic sampling algorithm applied to a single disorder realization on a random graph of size $N=100$ with $z=2$ ($\Delta=2$ and $H=0$). The black dots indicate the states in the $m$-$\mu$ space that are visited during the iteration procedure after (a) 1, (b) 20, and (c) 60 MC stages. The gray area is
    the domain of existence of all possible states.}
\end{figure}

The method is illustrated in Fig.~\ref{Fig2} which shows the exploration of the states in the $m$-$\mu$ plane at different MC stages.
In this case, the number of MC steps during the iteration procedure varied from $10^4$ initially to
typically $10^5$ during the last $\sim 15$ stages.  By definition, all the states
are contained in the domain ${\cal D}= \{(m,\mu)|m\in [-1,1],\mu\in[0,1]\}$.  During the first
stages (Fig.~\ref{Fig2}(a) and (b)), only the states close to $m=0$ and $\mu=0.5$ are visited. Those
close to the boundaries of ${\cal D}$ are less degenerate and are only visited after a
certain number of iterations (see Fig.~\ref{Fig2}(c)).  Note that certain regions of ${\cal D}$ are not visited 
because $m$ and $\mu$ are not independent quantities and the two coupled equations $m=(1/N)\sum_i s_i$ and $\mu = (1/N) \sum_i \Theta(-s_i f_i)$ (with $N$ unknowns $\{s_i\}$) may have no solution at all. When there are solutions, which correspond to the states visited by the algorithm, their number is precisely $\hat{\cal N}(m,\mu;H)$. 
Consider for instance the situation at the boundaries of
${\cal D}$.  For $m=\pm 1$, only one macrostate is visited (see
Fig.~\ref{Fig2}(c)).  Indeed, only the value $\mu=(1/N) \sum_i\Theta(\mp f_i)$ is compatible
with $m=\pm1$.  Therefore, $\hat{\cal N}(\pm1,\mu,H)=1$ for this value of $\mu$ and
zero otherwise.  Imposing $\mu=0$ or $\mu=1$ is less restrictive because
fixing $\mu$ only imposes some inequalities on the products $s_if_i$ (for instance, $\mu=0$
implies $s_if_i>0,\, \forall i$). Therefore, in general, several macrostates are visited on these two boundaries.

The metastable states, which correspond to $\mu=0$, are exponentially less degenerate than the
states with $\mu \sim 0.5$. Since the algorithm detects the most degenerate states
first, the enumeration of the metastable states becomes more and more difficult as the system size
increases. However, this method represents a noticeable improvement over the exact enumeration procedure and
we could study sizes up to $N=100$ (with $300$ disorder realizations).

\subsection{Simulated annealing}
\label{sec-SimAnn}
As introduced in Sec.~\ref{sec:Complex-Finite}, the magnetization $m_{\alpha}(g;H,N)$ which
corresponds to the maximum of $e^{Ngm} {\cal N}_{\alpha}(m;H,N)$ for a given disorder realization plays a
crucial role in our analysis of the distribution of the typical metastable states in the $H$-$m$
plane. In this section, we describe a Monte Carlo procedure that yields $m_{\alpha}(g;H,N)$ without exploring the whole
$(m,\mu)$ space. A similar procedure has been proposed in the context of granular materials \cite{Barrat2000} and spin systems \cite{Berg2001}.

By sampling the spin configurations with
the transition probability $p(\mathbf{s} \rightarrow
\mathbf{s'})=\min\{1,\exp(Ng(m'-m))\}$, one would obtain a distribution for the magnetizations which is proportional
to $e^{gNm} \hat{\cal N}(m,\mu;H,N)$.  According to the arguments of the previous section, the
maximum of this distribution is dominated by the most degenerate states which correspond
to $\mu>0$. This simple procedure is therefore inappropriate to analyze the statistics of the
metastable states ($\mu=0$). In order to sample properly these states, one needs to introduce a bias
parameter $g_{\mu}$ such that the transition from $\mathbf{s}$ to
$\mathbf{s'}$ is accepted with the probability $p(\mathbf{s} \rightarrow
\mathbf{s'})=\min\{1,\exp[N(A(m',\mu')-A(m,\mu))]\}$ where $A(m,g)=gm-g_{\mu}\mu$.  The interesting
limit for visiting the metastable states is then $g_{\mu} \rightarrow \infty$.  However, since these states are
much less degenerate than those with $\mu>0$, taking a large value of $g_{\mu}$ does not lead to the 
maximum of $e^{gNm} {\cal N}(m;H,N)$ right
away.  Instead, one needs to use a simulated annealing algorithm which consists in performing a sequence of MC runs, starting with $g_{\mu}=0$ and increasing slowly $g_{\mu}$ until no additional spin flip is accepted (we typically performed sequences of $2000$ MC runs, increasing $g_{\mu}$ by increments of $0.05$ up to $g_{\mu}=15$).

If ${\cal N}_{\alpha}(m;H,N)$ has a unique maximum, the magnetization resulting from this procedure is then  $m_{\alpha}(g;H,N)$. In practice, however, there may be several values of $m$ for which $e^{gNm} {\cal N}_{\alpha}(m;H,N)$ is close to its maximum and the final magnetization may slightly differ from the actual $m_{\alpha}(g;H,N)$.  Our simulations show that such deviations do exist but, as will be seen in Secs.~\ref{Sigma_Q-z2} and \ref{Sigma_Q-z4} by comparing with the 
analytical results for $z=2$ and the results of the other methods, they are not statistically relevant. The quenched complexity can then be obtained by considering $g$ as a function of $m$ and integrating numerically the Legendre equation $\partial \Sigma_Q / \partial m=-g$,

\begin{equation}
\label{Eq-SimAnn-1}
\Sigma_Q(m)=-\int_{m_{-\infty}^{}}^{m} g(m')dm'
\end{equation}
using the fact that $\Sigma_Q(m_{-\infty})=0$ since there is only one state along the hysteresis loop.

When ${\cal N}_{\alpha}(m;H,N)$ has two distinct maxima separated by a
gap (see e.g. Fig.~\ref{Fig9_5}) so that a threshold $m_{th}$ has to
be introduced in the analysis, the method does not seem
to work. Indeed, we found that convergence to the metastable states
(expected for $g_{\mu} \rightarrow \infty$) could not be achieved for all values of
$g$.  Nevertheless, it will be shown below that some useful informations
can be extracted from the data even if the two branches
$m^{\pm}(g;H,N)$ are not analyzed separately. With this algorithm,
systems with sizes up to $N=1000$ (with $300$ disorder realizations)
could then be investigated.

%%%%%%%%%%%%%%%%%%%%%%%%%%%%% z=2 %%%%%%%%%%%%%%%%%%%%%%%%
\section{Results for random graphs with $z=2$}
\label{z2}
We first present the numerical results for random graphs with connectivity $z=2$. In this case, the particular value of $\Delta$ does not introduce any qualitative change and we
arbitrarily take $\Delta=2$. Our main goal is to check the validity of the numerical approaches by
comparing to the analytical results of Ref.~\onlinecite{Detcheverry2005} and testing whether the different
methods are consistent with each other.

\subsection{Annealed complexity}
\label{Sigma_A-z2}
\begin{figure}
\epsfig{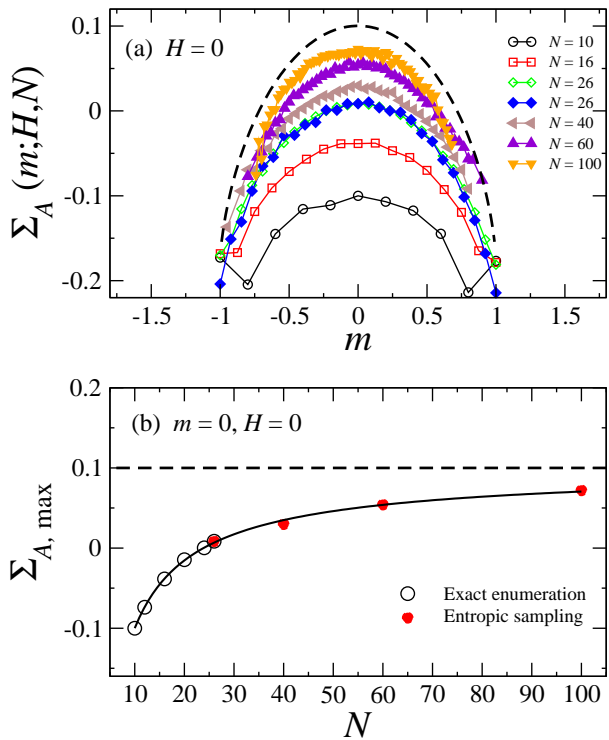}
\caption{\label{Fig3} (a) Annealed complexity $\Sigma_A(m;H,N)$ for $H=0$ and various system sizes (random graphs with $z=2$
  and $\Delta=2$). Open symbols: exact enumeration; solid symbols: entropic sampling. The dashed line corresponds to the analytical result of Ref.~\onlinecite{Detcheverry2005}. (b) Dependence of $\Sigma_{A,\max}$ on the system size $N$. The solid line is a fit to the data. The horizontal dashed line indicates the exact value $\Sigma_{A,\max}^{\infty}=0.1$ calculated in Ref.~\onlinecite{Detcheverry2005}.(Color on line.)}
\end{figure}
We first focus on the annealed complexity as defined by Eq.~(\ref{Complex.Eq.1}). 
Fig.~\ref{Fig3}(a) shows $\Sigma_A$ as a function of $m$ at $H=0$ for different sizes $N$.  The data for $N \leq 26$  have been obtained by exact enumeration and those for $N \geq 26$ by entropic sampling. (Recall that the simulated
annealing method is only appropriate to compute the typical quantities.) The good overlap of the two sets of data for $N=26$ shows that the two methods are consistent. At fixed $m$, the complexity increases with $N$ with a clear tendency towards the value $\Sigma_A^{\infty}(m;H=0)$ computed in Ref.~\onlinecite{Detcheverry2005} (dashed line in the figure). Let us analyze in
more detail the case $m=0$ that corresponds to the maximum of the complexity at zero field. The size-dependence of $\Sigma_{A,\max}$ plotted in Fig.~\ref{Fig3}(b) is well described
by the ansatz $\Sigma_{A,\max}=\Sigma_{A,\max}^{\infty}+aN^{-1}\ln N+bN^{-1}$ involving the three
free parameters $\Sigma_{A,\max}^{\infty}$, $a$, and $b$\cite{note2}.  A least-squares fit to the data yields
$\Sigma_{A,\max}^{\infty}=0.10$, $a=-0.46$, $b=-0.94$, and the value of $\Sigma_{A,\max}^{\infty}$
is in remarkable agreement with the analytical result of Ref.~\onlinecite{Detcheverry2005} (actually, a fit of the data for $N \leq 26$ already provides a good estimate of $\Sigma_{A,\max}^{\infty}$). One can perform a similar analysis of the data for $m\neq 0$ but, because of the
discreteness of the overall magnetization $M$, an interpolation of the curves
$\Sigma_A(m;H,N)$ is then necessary to study the dependence with $N$ at fixed $m$.

\begin{figure}
\epsfig{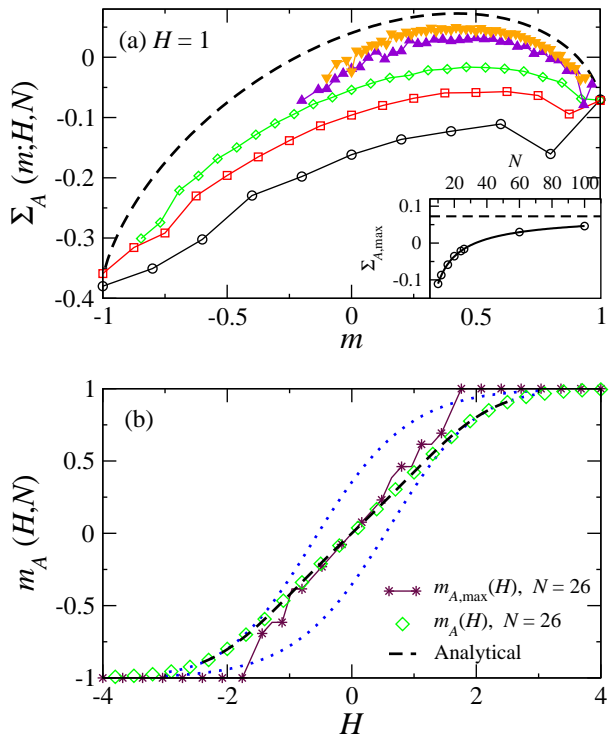}
\caption{\label{Fig4} (a) Annealed complexity $\Sigma_A(m;H,N)$ for $H=1$ and various system sizes (random graphs with
  $z=2$ and $\Delta=2$). The convention for symbols and lines is the same as in Fig.~\ref{Fig3}. The
  inset shows the size dependence of the complexity at the maximum.  (b) Average magnetization $m_A(H)$ obtained from Eq. (\ref{Eq-z2-mA}) (diamonds) and magnetization $m_{A,\max}(H,N)$ at the maximum of $\Sigma_A(m)$ (stars) for $N=26$.  The dashed line  is the analytical result of Ref.~\onlinecite{Detcheverry2005} and the dotted line represents the saturation hysteresis loop calculated in Ref.~\onlinecite{Dhar1997}.(Color on line.)}
\end{figure}

We now consider the complexity at $H=1$ (see Fig.~\ref{Fig4}(a)) to illustrate the performance of our numerical procedures in non-zero field. The curves in Fig.~\ref{Fig4}(a) are now asymmetric, with a maximum at a positive magnetization.  Note that the entropic sampling method is not able to explore the whole range of magnetizations in the largest systems. Indeed, when the field $H$ is positive, the states with $m<0$ are much less degenerate than those with $m>0$, and are therefore more difficult to detect. For
instance, using the value of the complexity for $N=100$, the probability of visiting a state with
$m=-0.2$ is approximately 5000 times smaller than the probability of visiting the states with
maximal complexity (with $m\simeq 0.42$). The detection of rare
states is also limited by the number of disorder realizations investigated. This number is only $\sim 200$ for large systems, which is significantly smaller than the $5000$ realizations generated in the exact enumeration. However, we shall
show below that the states which are not detected in large systems are nontypical and do not
affect the quenched averages.

The numerical data for finite systems again underestimate $\Sigma_{A}^{\infty}(m,H)$ and the results must be extrapolated to $N\rightarrow \infty$. The dependence of the 
maximal complexity $\Sigma_{A,\max}$ with $N$ is shown in the inset of Fig.~\ref{Fig4}. The same fit as
for $H=0$ yields $\Sigma_{A,\max}^{\infty}=0.079$, which is in reasonable agreement with the exact value $0.073$. 

The curve $m_{A,\max}(H)$ shown in Fig.~\ref{Fig4}(b) represents the locus of the maxima of the
annealed complexity in the $H$-$m$ plane. In the thermodynamic limit, the (annealed) average magnetization of the states defined for finite systems as
\begin{equation}
\label{Eq-z2-mA}
 m_A(H,N)=\frac{\sum_m m \overline{{\cal N}(m;H,N)}}{\sum_m \overline{{\cal N}(m;H,N)}}
\end{equation}
is exponentially dominated by $m_{A,\max}^{\infty}(H)$ and therefore
$m_A^{\infty}(H)=m_{A,\max}^{\infty}(H)$. In small systems, however, $m_A(H,N)$  coincides with $m_{A,\max}(H,N)$ at small fields only. In particular, $m_{A,\max}(H,N)$ reaches saturation before $m_A(H,N)$. The deviations decrease with system size and the
two curves are expected to coincide for large $N$.  Interestingly, finite-size effects on
$m_A(H,N)$ are much less important than on $m_{A,\max}(H,N)$, and $m_A(H,N)$ matches the analytical prediction for $m_A^{\infty}(H)$ even for small $N$.

\subsection{Quenched complexity}
\label{Sigma_Q-z2}
\begin{figure}
\epsfig{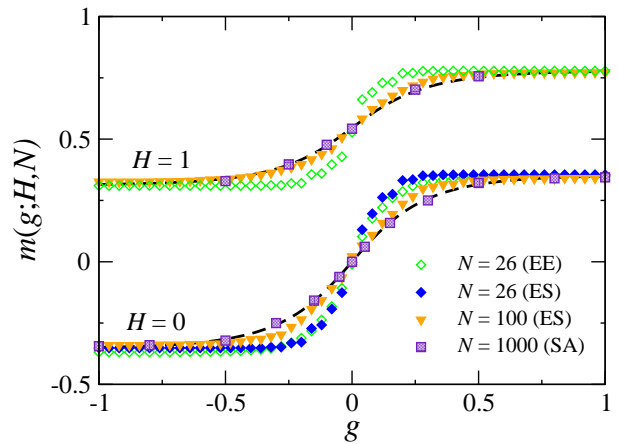}
\caption{\label{Fig5} Magnetization $m(g)$ for $H=0$ and $H=1$, and various system sizes (random graphs with $z=2$ and $\Delta=2$). Open, solid, and shaded symbols correspond to the data obtained from exact enumeration (EE), entropic sampling (ES), and simulated annealing (SA), respectively. The dashed lines represent the analytical results of Ref.~\onlinecite{Detcheverry2005}.(Color on line.)}
\end{figure}
\begin{figure}
\epsfig{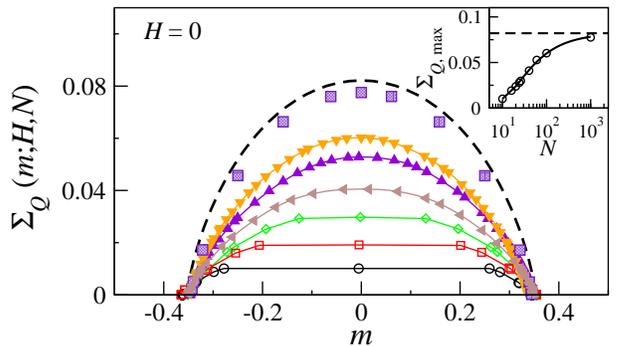}
\caption{\label{Fig6} (a) Quenched complexity $\Sigma_Q(m)$ at $H=0$ for various system sizes (random graphs with $z=2$ and $\Delta=2$). The large squares correspond to $N=1000$ and the other symbols and lines refer to the same sizes as in Fig.~\ref{Fig3}. The dashed line is the analytical result for $N\rightarrow \infty$. The inset
  shows the size dependence of $\Sigma_{Q,\max}$, the maximal value of the complexity.  The solid
  line is a fit to the data.  The horizontal dashed line indicates the theoretical value
  $\Sigma_{Q,\max}^{\infty}=0.082$ calculated in Ref.~\onlinecite{Detcheverry2005}.(Color on line.)}
\end{figure}

According to the analytical results of Ref.~\onlinecite{Detcheverry2005},
the quenched complexity for $z=2$ displays a single maximum as a
function of $m$. Therefore, the numerical computations can be based on
the definition of $\Sigma_Q$ given in Eq.~(\ref{Complex.Eq.4}). The
first test is to check if the different numerical methods provide a
good estimate of $m(g;H,N)$. In Fig.~\ref{Fig5}, we plot $m(g;H=0,N)$
and $m(g;H=1,N)$. For $N=26$, we compare the results obtained by exact
enumeration and by entropic sampling, and the good agreement between the two curves 
shows that the two methods are consistent. The
results for $N=1000$ only correspond to the simulated annealing
method.  The deviations of the data from the analytical
result of Ref.~\onlinecite{Detcheverry2005} are due to finite-size effects,
and the agreement becomes very good for $N=1000$. Even more important
is the fact that the curves converge towards the correct limit when
$g\rightarrow \pm \infty$ (actually, $m(g;H,N)$ is essentially
constant when $\mid g\mid \gtrsim 1$) in spite of the lack of data for
negative magnetizations in large systems (see Fig.~\ref{Fig4}(a)).
Such a result confirms that (i) the states with negative $m$ which are
not detected by the entropic sampling method in large systems are not typical, and (ii) the range of
magnetizations for which the density of the typical states is finite
(i.e., $\Sigma_Q>0$) is well captured by the numerical methods.

Fig.~\ref{Fig6}(a) shows the quenched complexity $\Sigma_Q(m)$ for
$H=0$, computed from Eqs.~(\ref{Complex.Eq.3b}) and
(\ref{Complex.Eq.4}), or, for $N=1000$ (with the simulated annealing
method), from Eq.~(\ref{Eq-SimAnn-1}). As could be anticipated from
the preceding figure, the range of magnetizations for which
$\Sigma_Q>0$ does not depend significantly on $N$. In contrast, the
values of $\Sigma_Q(m)$ in this interval display significant
finite-size effects. As for the annealed case, the actual complexity
in the thermodynamic limit is underestimated but the curves
extrapolates to the correct limit. For instance, the extrapolation of
the maximal complexity yields $\Sigma_{Q,\max} ^{\infty}=0.082$ which
is in excellent agreement with the value computed in
Ref.~\onlinecite{Detcheverry2005}.

\section{Results for random graphs with $z=4$}
\label{z4}
\begin{figure}
\epsfig{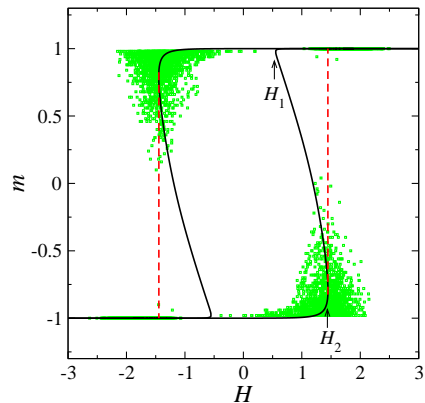}
\caption{\label{Fig7} Saturation hysteresis loop on random graphs with $z=4$ and $\Delta=1$. The solid line is
  the curve obtained in the thermodynamic limit from the equations of Ref.~\onlinecite{Dhar1997}. The intermediate ``unstable'' parts (in the range $H_1<H<H_2$ for the ascending branch) are included. The hysteresis loop has a jump
 discontinuity at $H=\pm H_2$ (dashed lines). The symbols represent the loops obtained numerically in 
  1000 disorder realizations of size $N=100$ (for clarity, the points are not connected).(Color on line.)}
\end{figure}

In this section devoted to random graphs with $z=4$, we shall only
focus on the features of the metastable states in the low disorder
regime ($\Delta < \Delta_c$) for which the hysteresis loop exhibits a
jump discontinuity, as displayed in Fig.~\ref{Fig7}. The behavior for
$\Delta > \Delta_c$ is indeed qualitatively similar to the one
discussed above for $z=2$. Specifically, we shall take $\Delta=1$
which is smaller than the critical value
$\Delta_c=1.78126$\cite{Dhar1997}.  It has been shown that the jump in
magnetization corresponds mathematically to the appearance of a new
stable root in the self-consistent equation for the fixed-point
probability derived in Ref.~\onlinecite{Dhar1997}. This equation has three
real solutions in the range $H_1<H<H_2$ (hereafter, for simplicity, we
only consider the lower branch since the upper one can be obtained by
symmetry), with two of them merging and becoming complex at $H_1$ and
$H_2$. As discussed in Ref.~\onlinecite{Detcheverry2005}, it is tempting to
associate the reentrant curve corresponding to the ``unphysical'' root
of the equation in the range $H_1<H<H_2$ to the limit of existence of
the metastable states, i.e., to the contour $\Sigma_Q(m,H)=0$.

However, before discussing this issue which concerns the quenched quantities (the only
ones that are actually observable), let us first summarize the results of our numerical analysis for
the annealed quantities. This will provide an additional comparison with the analytical results of
Ref.~\onlinecite{Detcheverry2005}.

\subsection{Annealed complexity}
\label{Sigma_A-z4}

\begin{figure}
\epsfig{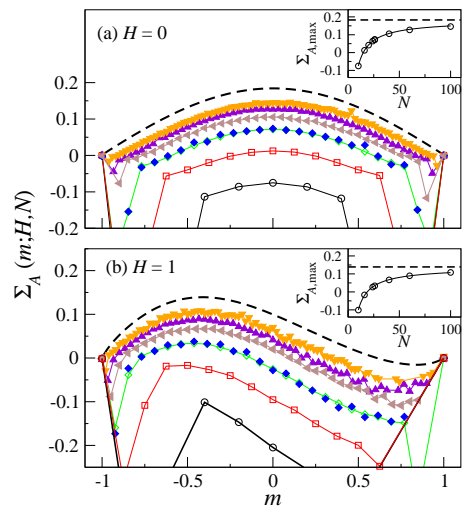}
\caption{\label{Fig8} Annealed complexity $\Sigma_A(m;H,N)$ for $H=0$ (a) and $H=1$ (b) and various system sizes (random graphs with $z=4$ and $\Delta=1$). The convention for symbols and lines is the same as in
  Fig.~\ref{Fig3}. The insets show the size dependence of the maximal complexity $\Sigma_{A,\max}$
  and the corresponding fits.(Color on line.)}
\end{figure}

The behavior of the annealed complexity in zero field is displayed in Fig.~\ref{Fig8}(a) for different
system sizes. It is similar to the one observed for $z=2$ and the function remains concave in
the whole range $-1\leq m \leq 1$. Moreover, the results nicely extrapolates to the thermodynamic
limit. For instance, as shown in the inset of Fig.~\ref{Fig8}(a), $\Sigma_{A,\max}$, the complexity at the 
maximum, extrapolates to the value $\Sigma_{A,\max}^{\infty}=0.183$ which is in excellent
agreement with the theoretical value of Ref.~\onlinecite{Detcheverry2005}.

The behavior in non-zero fields is qualitatively different. In this case, the complexity is not
concave in the whole range of magnetizations and two local maxima may be present. This is
illustrated in Fig.~\ref{Fig8}(b) for $H=1$. In this case, the global maximum of the complexity is
located at negative magnetizations although the external field is positive. This is the general
behavior observed at small values of $H$. When the field is further increased, the peak at positive
magnetizations (which is very close to $m=1$ for this value of $\Delta$) becomes the global maximum. As will be
shown below, the quenched complexity displays a similar behavior.

The numerical results for finite systems again nicely extrapolate to the exact thermodynamic limit
computed in Ref.~\onlinecite{Detcheverry2005}. The extrapolated result for $ \Sigma_{A,\max}$ is
$0.139$ (inset in Fig.~\ref{Fig8}(b)), which is in very good agreement with the theoretical value
$\Sigma_{A,\max}^{\infty}=0.138$. One may note that the corresponding magnetization only deviates from its limiting value $m_{A,\max}^{\infty}=-0.41$ by amounts smaller than
$2/N$.

\subsection{Quenched complexity}
\label{Sigma_Q-z4}

Our study of the quenched average is based on the data obtained by entropic sampling and simulated annealing on
random graphs of size $N=100$ and $N=1000$. The hysteresis loops obtained in $1000$
realizations of the disorder are shown in Fig.~\ref{Fig7} and illustrate the typical behavior of the magnetization curve in finite systems (note in particular that the average value of the field corresponding to the jump in $m$ is $H=1.72$ whereas the coercive field in the thermodynamic limit is $H=1.45$).

\begin{figure}
\epsfig{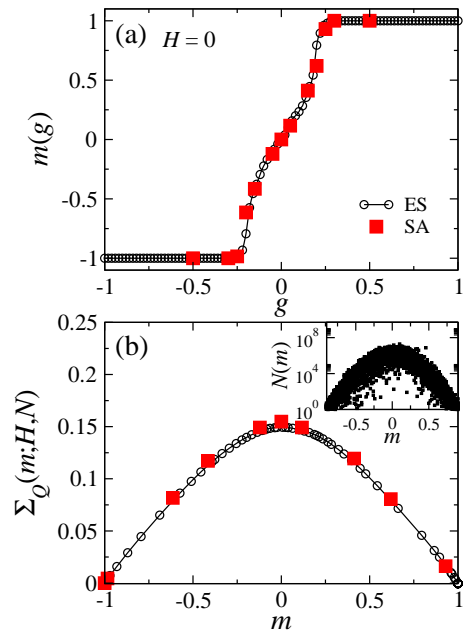}
\caption{\label{Fig9} (a) $m(g)$ and (b) $\Sigma_Q(m;H,N)$ for $H=0$ (random graphs with $z=4,N=100$, and $\Delta=1$). Circles and squares correspond respectively to the entropic sampling and the simulated annealing method.  In (b), the inset displays the raw data for ${\cal N}(m;H=0,N)$ (obtained with the entropic sampling method) in a semi-log scale.(Color on line.)}
\end{figure}

Let us first consider the case $H=0$. As shown in the inset of Fig.~\ref{Fig9}(b), the number of metastable states ${\cal N}(m;H=0,N)$ has a single maximum at $m \approx 0$, which implies that one can use a single Legendre transform to compute $m(g)$ and $\Sigma_Q(m)$ 
(Eqs.(\ref{Complex.Eq.2})-(\ref{Complex.Eq.4})). As can be seen in Fig.~\ref{Fig9}(a), $m(g)$ increases monotonically from $m_{-\infty}\approx-1$ to
$m_{+\infty}\approx+1$ (note the good agreement between the entropic sampling and simulated annealing methods).These two values, which are extremely close (but not equal) to $\pm 1$, represent the lower and upper bounds of the magnetization of the typical
states and correspond to the lower and upper branches of the hysteresis
loop, respectively (one can see in Fig.~\ref{Fig7} that these values are indeed very close to $\pm
1$).  The corresponding behavior of the complexity is shown in Fig.~\ref{Fig9}(b). The curve is
of course symmetric with respect to $m=0$ (one cannot see on the scale of the figure that 
the slope $\partial \Sigma_Q(m)/ \partial m$ is infinite at $m=m_{\pm \infty}$).

Although the method based on two distinct Legendre transforms is neither
appropriate nor necessary in this case, one may check that it yields the same results and that there is no gap in magnetization for any choice of $m_{th}$ in the interval $[m_{-\infty},m_{+\infty}]$. According to the arguments
in Sec.~\ref{sec:Complex-Finite}, this shows that the density of the typical states is strictly
positive when $m_{-\infty}\le m \le m_{+\infty}$, that is to say inside the hysteresis loop.

\begin{figure}
\epsfig{file=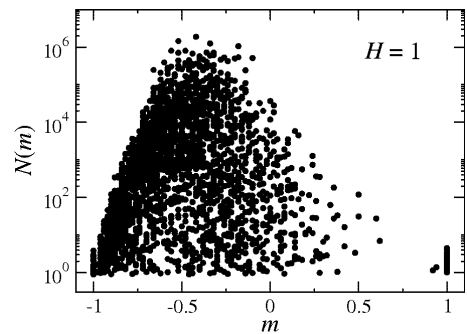, width=6cm,clip=}
\caption{\label{Fig9_5} Number of metastable states
  ${\cal N}(m;H=1,N)$ obtained by entropic sampling in $200$ disorder realizations (random graphs with $z=4$, $N=100$, and $\Delta=1$). Note that the right lobe is concentrated very close to $m=1$.}
\end{figure}

We next consider the case $H\neq 0$ and focus on the interesting region around the jump in magnetization. As can be seen in Fig.~\ref{Fig9_5} that shows the raw data for ${\cal N}(m;H=1,N=100)$ obtained in $200$ samples, it is now crucial to introduce two Legendre functions
$\Lambda^{-}(g)$ and $\Lambda^{+}(g)$ in order to take into account the possible existence of a gap in the magnetization of the typical states, which would imply that the curve $\Sigma_Q(m,H)=0$ has a reentrant part, as anticipated in
Fig.~\ref{Figintro}(b). Whether or not this part of the curve is described by the ``unphysical'' root of the
fixed-point equation of Ref.~\onlinecite{Dhar1997} (corresponding to the branch between $H_1$ and $H_2$ drawn in Fig.~\ref{Fig7}) is an independent issue which is discussed below.

\begin{figure}
\epsfig{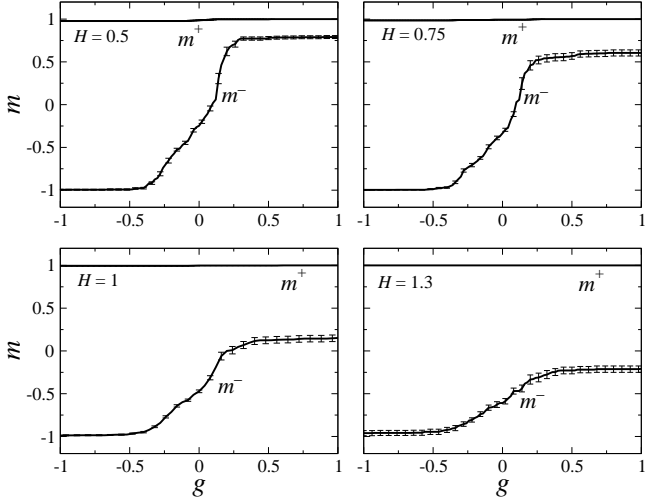}
\caption{\label{Fig10} Magnetizations $m^{-}(g)$ and $m^{+}(g)$ for different external fields chosen in the vicinity 
of the jump discontinuity
  (random graphs with $z=4$, $N=100$, and $\Delta=1$). The threshold in magnetization is $m_{th}=0.9$
  in all cases.}
\end{figure}

Fig.~\ref{Fig10} summarizes the results for the magnetizations $m^{\pm}(g)$ obtained by entropic sampling. In contrast with the case $H=0$, we find
that $m^{-}_{+\infty}$ is strictly smaller than $m^{+}_{-\infty}$ for the four values of $H$
considered in the figure (note that $m^{+}(g)$ is always very close to 1). This points towards the existence of a gap and two separate branches $m^{\pm}(g)$ in the thermodynamic limit. Note that these curves have
been calculated with the threshold value $m_{th}=0.9$. The results do not depend on 
$m_{th}$, however, as long as the value is chosen in an appropriate interval whose length increases with $H$.  For instance, for $H=0.5$ and $H=1.3$, the approximate intervals are $[0.85,0.95]$
and $[0.2,0.95]$, respectively. Outside these intervals one gets
$m^{-}_{+\infty}=m^{+}_{-\infty}$, which simply indicates that there is a finite density of typical
states with magnetization $m_{th}$.

\begin{figure}
\epsfig{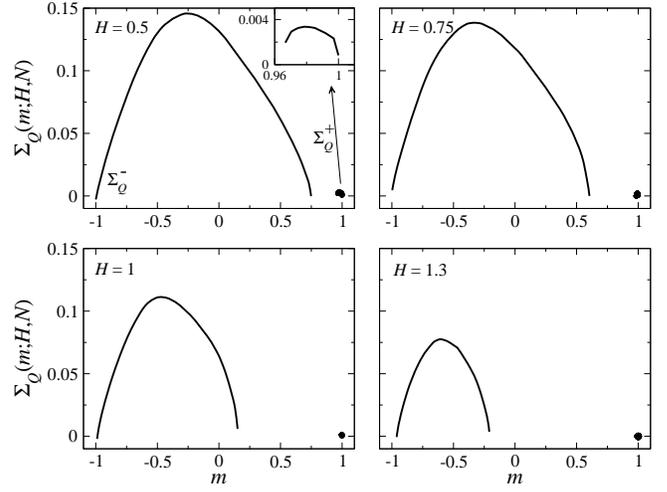}
\caption{\label{Fig11} Quenched complexity $\Sigma_Q(m;H,N)$ for $H=0.5,0.75,1$ and $1.3$ (random graphs
  with $z=4$, $N=100$, and $\Delta=1$). The two branches correspond to $\Sigma_Q^{-}(m)$ and $\Sigma_Q^{+}(m)$, respectively. For $H=0.5$, the inset shows a magnification of $\Sigma^{+}(m)$. (The threshold in magnetization
  is $m_{th}=0.9$ in all cases.)}
\end{figure}

The corresponding complexities $\Sigma_Q^-$ and $\Sigma_Q^+$ computed
from Eqs.~(\ref{Complex.Eq.6}) are shown in Fig.~\ref{Fig11}. The
two-peak structure resembles that of the annealed complexity in
Fig.~\ref{Fig8}(b), but whereas $\Sigma_A$ has a minimum between the
two maxima, $\Sigma_Q$ is not even defined in the interval
$m^{-}_{+\infty}<m<m^{+}_{-\infty}$. This gap, whose size increases as
$H$ approaches the coercive field, indicates that there is no typical
metastable state having a magnetization in this range. (Note that
$\Sigma_Q^-$ and $\Sigma_Q^+$ seem to go to zero at
$m=m^{-}_{+\infty}$ and $m=m^{+}_{-\infty}$, respectively. It is
possible, however, that the complexity is small but finite at these
borders, in contrast with what happens for $m=m^{-}_{-\infty}$ and
$m=m^{+}_{+\infty}$, i.e., along the two branches of the hysteresis
loop.)

In table~\ref{Table-mg-P}, the results for $m^{-}_{-\infty}$ and $m^{-}_{+\infty}$ are compared
to the values of the magnetization obtained from the solutions of
the fixed-point equation of Ref.~\onlinecite{Dhar1997} (the other branch $\Sigma_Q^{+}(m)$ is so close to $m=+1$ that 
comparing the two limits $m^{+}_{-\infty}$ and $m^{+}_{+\infty}$ with the theoretical values is not meaningful). Whereas
$m^{-}_{-\infty}$ is in reasonable agreement with the theoretical value for the lower branch of the hysteresis loop, $m^{-}_{+\infty}$ significantly deviates from the solution that describes the
``unstable'' branch drawn in Fig.~\ref{Fig7}. This may be due, of course, to finite-size effects. In particular,
the fact that our data predicts a discontinuity in the magnetization for $H=0.5$ whereas the
fixed-point equation of Ref.~\onlinecite{Dhar1997} has only one root at this field is somewhat surprising.
However, this may also indicate that the ``unstable'' branch obtained analytically does not represent the limit of existence of the metastable states in the low disorder regime. As far as we know, there is indeed no proof that the two curves should coincide.

\begin{table}[h]
%\begin{landscape}
\begin{center}
\begin{tabular}{ccccc}
\hline
\hline
$H$ & $m_{hyst}^-$ & $m^{-}_{-\infty}$ & $m_{unst}$ & $m^{-}_{+\infty}$\\
\hline
0.5 & -0.999 & $-0.995 \pm 0.005$ & - & $0.79 \pm 0.02$ \\
0.75 & -0.998  & $-0.997 \pm  0.002$ & $0.7$ & $0.61 \pm 0.04$\\ 
1 & -0.994 & $-0.991 \pm 0.004$ & $0.33$ & $0.17 \pm 0.05$\\
1.3 & -0.97 & $-0.96 \pm 0.03$ & -0.24 & $-0.21 \pm 0.04$\\
\hline
\hline
\end{tabular}

\caption{Comparison of $m^{-}_{-\infty}(H,N)$ and $ m^{-}_{+\infty}(H,N)$ with the values of the magnetization obtained from the solution of the fixed-point equation of Ref.~\onlinecite{Dhar1997} (random graphs with $z=4, N=100$ and $\Delta=1$). $m_{hyst}^-$ and $m_{unst}$ are the theoretical values corresponding to the ascending branch of the hysteresis loop and the ``unstable'' intermediate branch, respectively (for $H=0.5$, the unstable solution does not exist).}
\label{Table-mg-P}
\end{center}
%\end{landscape}
\end{table}

\begin{figure}
\epsfig{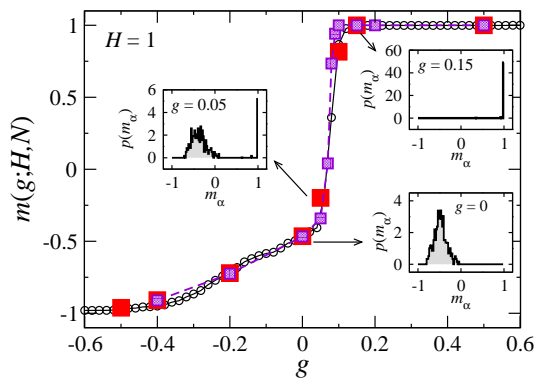}
\caption{\label{Fig12} Magnetization $m(g)$ for $H=1$ obtained from a single Legendre transform (random graphs with $z=4$ and $\Delta=1$). Circles correspond to the entropic sampling method ($N=100$)
  and squares to the simulated annealing method ($N=100$: large squares, $N=1000$: small squares). The insets
  show the normalized distribution of $m_{\alpha}(g)$ obtained by simulated annealing for
  $g=0,0.05,0.15$ and $N=100$. (Color on line.)}
\end{figure}

In order to study larger systems and check that the above conclusions
are correct, one has to resort to the simulated annealing procedure
(this is mandatory in the case of the cubic lattice, as discussed
below in section \ref{Cubic-Lattices}). Unfortunately, as noted in section
\ref{sec-SimAnn}, this method does not work properly when one introduces a threshold
$m_{th}$ so to take into account the fact that ${\cal
  N}(m;H,N)$ has two maxima. In consequence, we use one Legendre transform only, which
has the unpleasant consequence that the two branches $m^{\pm}(g)$ are
mixed. Nevertheless, as shown in Fig.~\ref{Fig12} for $H=1$, the
existence of a gap in the magnetization of the metastable states has a
clear signature in $m(g)$. In order to compare with the entropic
sampling method, let us first consider the case of random graphs of
size $N=100$. One can see that the two methods are in perfect
agreement. The comparison with the results obtained with two Legendre
transforms (see Fig.~\ref{Fig10}) shows that $m(g)=m^{-}(g)$ for
$g<g_1 \simeq 0.04$ and $m(g)=m^{+}(g)$ for $g>g_2 \simeq 0.15$. In
the range $g_1<g<g_2$, the magnetization varies continuously but
steeply from the branch $m^{-}(g)$ to the branch $m^{+}(g)$. This
behavior comes from the fact that the distribution $p(m_{\alpha})$ of
the magnetizations $m_{\alpha}(g)$ has a single peak for $g<g_1$
(because $\Lambda_{\alpha}^{-}(g)>\Lambda_{\alpha}^+(g)$ in all
disorder realizations) or $g>g_2$ (because
$\Lambda_{\alpha}^{-}(g)<\Lambda_{\alpha}^+(g)$), and two peaks for
$g_1<g<g_2$ (because there are some realizations for which
$\Lambda_{\alpha}^{-}(g) < \Lambda_{\alpha}^+(g)$). Therefore, in the
latter case, the ``mixed'' quantity $m(g)$ does not represent properly
the distribution of the magnetizations. However, the rapid increase of
$m(g)$ signals unambiguously the passage from one branch to the other
one (compare for instance with the case $z=2$ in Fig.~\ref{Fig5}). One
can see that the slope becomes steeper as one goes from $N=100$ to
$N=1000$, and, therefore, one expects a 
discontinuity in the thermodynamic limit. We stress, however, that this
discontinuity has no physical meaning and that two branches $m^{\pm}(g)$ coexist from $g=-\infty$ to $+\infty$ for
this value of $H$.

\begin{figure}
\epsfig{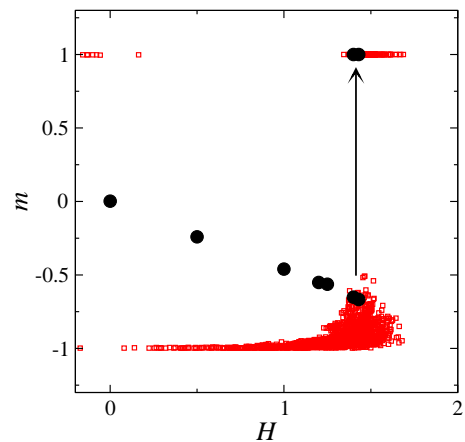}
\caption{\label{Fig13} Typical magnetization $m(g=0;H)$ (solid circles) of the metastable states for $H>0$ (random graphs with $z=4$, $N=1000$, and $\Delta=1$). The data are obtained by the simulated annealing method. 
 The small squares represent the hysteresis loops obtained in $100$ disorder realizations (for clarity, the points are not connected). The case $H<0$ is obtained by symmetry. (Color on line.)}
\end{figure}

Finally, in Fig.~\ref{Fig13}, we show the curve $m(g=0;H)$ which
represents the locus of the maximum of the quenched complexity in the
$H-m$ plane and corresponds to the typical magnetization of the
metastable states. As could be seen already in Fig.~\ref{Fig11}, the
maximum for $H \lesssim 1.3$ is located at negative magnetizations
despite the fact that the external field is positive. This behavior is
somewhat unexpected (see the discussion in section 5 of
Ref.~\onlinecite{Detcheverry2005}) and similar to that of the (annealed)
average magnetization $m_A(H)$ (Fig. 9 in
Ref.~\onlinecite{Detcheverry2005}). This is in contrast with the behavior
observed for $z=2$ and for $z=4$ in the strong disorder regime
(although the change of behavior does not exactly occur at
$\Delta=\Delta_c$). When $H$ is further increased, the maximum of
$\Sigma_Q^+$ becomes larger than the maximum of $\Sigma_Q^-$ and
$m(g=0;H)$ jumps to a positive value very close to $1$ (the arrow in
the figure indicates the approximate average field at which the two
maxima are equal in systems of size $N=1000$). As we shall see in the
following section, the qualitative behavior on the cubic lattice is
quite similar.

\section{Results for the cubic lattice}
\label{Cubic-Lattices}
\begin{figure}[h]
\epsfig{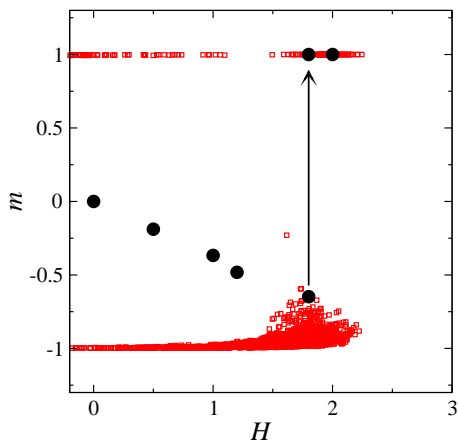}
\caption{\label{Fig15} Saturation hysteresis loops (small squares) for $100$ disorder realizations on a cubic lattice 
with $N=10^3$ and $\Delta=1.8$ (for clarity, the points are not connected). The solid circles represent the typical magnetization $m(g=0,H)$ for $H>0$ obtained by simulated annealing.(Color on line.)}
\end{figure}
In this section, we present some results for the metastable states of the RFIM on the cubic lattice. 
We focus on the low disorder regime where the hysteresis loop is discontinuous. 
Our only objective  is to show that there exists a gap in magnetization with 
no metastable states for a certain value of $H$ smaller than the coercive field. By continuity, this implies that there is a whole region inside the hysteresis loop where the number of states is zero. Since the entropic sampling method is limited to systems which are too small (say, $N <5\times5\times5$), we have used the simulated annealing algorithm and analyzed the results with a single Legendre transform, as discussed above. The system size is $N=10\times10\times10$ with typically 100 to 300 disorder realizations. We have chosen $\Delta=1.8$ which is smaller than the critical disorder $\Delta_c \simeq 2.2$ estimated in Refs.\cite{Perkovic1999,PerezReche2003}). 
\begin{figure}[h]
\epsfig{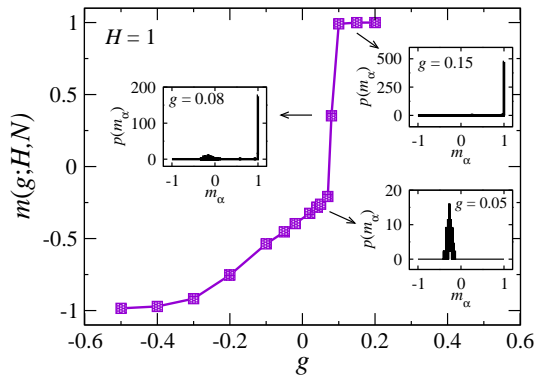}
\caption{\label{Fig16}  Magnetization $m(g)$ for $H=1$ obtained from a single Legendre transform (simulated annealing method on a cubic lattice with $N=1000$ and $\Delta=1.8$).  The insets
  display the normalized distribution of $m_{\alpha}$ for $g=0.05,0.08$ and $0.15$.(Color on line.)}
\end{figure}

In Fig.~\ref{Fig15}, we show the hysteresis loops obtained in 100
disorder realizations. All of them display a big jump in magnetization
at a coercive field whose average value is $H=1.95$.  For $H=1$, the
behavior of the magnetization $m(g)$ is quite similar to the one
observed on random graphs for $z=4$. As shown in Fig.~\ref{Fig16},
there is a steep increase in $m(g)$ around $g \approx 0.08$ (to be
contrasted with the 1d case shown in Fig.~\ref{Fig5} or the situation
for $\Delta >\Delta_c$), with a displacement of the peak in
$p(m_{\alpha})$, the distribution of the magnetizations
$m_{\alpha}(g;H,N)$, from $m\approx -0.25$ to $m\approx 1$. The
interval in which two peaks are present simultaneously is very small.
This suggests that there are in fact two branches $m^{\pm}(g)$, each
one corresponding to a distinct branch $\Sigma^+(m)$ or $\Sigma^-(m)$
of the quenched complexity as in Fig.~\ref{Fig11}.  The similarity with the
scenario on random graphs is also confirmed by the behavior of
$m(g=0;H)$, the typical magnetization of the metastable states. In
Fig.~\ref{Fig15}, like in Fig. \ref{Fig13}, the typical magnetization first
decreases as the field increases (revealing that the maximum of the
complexity occurs at negative magnetizations), and finally, at a field
that is smaller than the average coercive field, jumps to a value very
close to $+1$.

Note that the fact that the domain of existence of the metastable states (in the $H-m$ plane) for the cubic lattice is nonconvex is also suggested by the local mean-field calculations done in Ref.~\onlinecite{Illa2006a} in which the system is driven by the magnetization at finite temperature (see Fig. 5 in that reference).

\section{Summary and conclusions}
\label{Conclusion}

We have studied numerically the distribution of the 1-spin-flip stable states in the RFIM at $T=0$ on random regular graphs and on the cubic lattice. Three complementary numerical methods (exact enumeration, entropic sampling, and simulated annealing) have been used to calculate the annealed and quenched complexities as a function of the magnetization and the field. The three methods have been shown to yield consistent results when applied simultaneously to the same sample.
The entropic sampling algorithm allows one to study significantly larger systems than exact enumeration while obtaining interesting informations on the whole configurational space\cite{note3}. The simulated annealing algorithm, however, is the only one which is adapted to large systems.

In the case of random graphs, comparison with the analytical results of Ref.~\onlinecite{Detcheverry2005} (for $\Sigma_A(m,H)$ and $\Sigma_Q(m,H)$ when $z=2$ and for $\Sigma_A(m,H)$ when $z=4$) shows that the numerical data can be successfully extrapolated to the thermodynamic limit. In particular, for $z=4$, the concave character of $\Sigma_A(m)$ for $H=0$ and the non-concavity for $H\neq 0$ are well captured by the numerical results. 

The present study allows us to confirm that in the low disorder regime (i.e., for $\Delta<\Delta_c$ on random graphs with $z=4$ or on the cubic lattice), there is a whole region inside the hysteresis loop without any typical metastable states, as illustrated in Fig. \ref{Figintro}(b). In this case, the quenched complexity $\Sigma_Q(m;H)$ is positive in two distinct regions separated by a finite gap in magnetization (for $H$ close but smaller than the coercive field). The existence of the two branches is also confirmed  by the discontinuous behavior of the typical magnetization of the states as a function of the field. In other words, a macroscopic discontinuity along the hysteresis loop merely indicates that there is a ``hole'' in the distribution of the metastable states (i.e. the domain of existence of the metastable states in the $H-m$ plane is nonconvex). It is likely that this explanation is valid for all driven systems at $T=0$ (at least when the hysteresis loop is the outer bound of all metastable states, which is the case when the coupling interactions are ferromagnetic).

\begin{figure}[h]
\epsfig{file=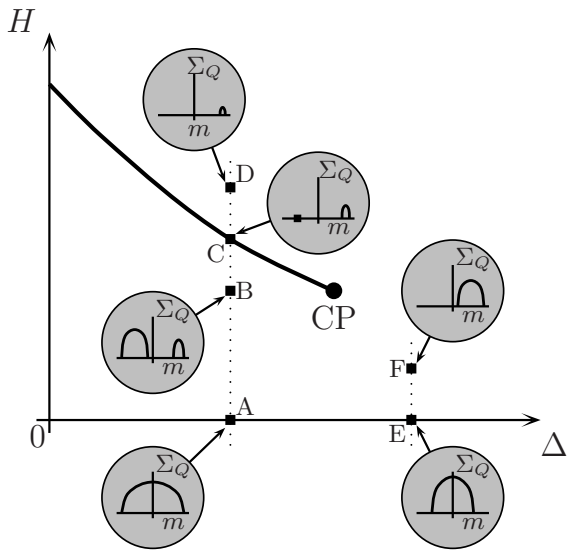, width=7.5cm,clip=}
\caption{\label{Fig20} Schematic behavior of the quenched complexity $\Sigma_Q(m)$ in the $\Delta$-$H$ plane
for different values of $\Delta$ and $H$ in the case where a disorder-induced phase transition exists. The solid line 
represents the coercive field $H_{coer}(\Delta)$ at which the
macroscopic discontinuity in the magnetization occurs. This line ends at the critical point (CP). Points 
A to D correspond to the low disorder regime ($\Delta<\Delta_c$) and points E and F to the strong-disorder regime 
($\Delta>\Delta_c$). The diagram for $H<0$ is obtained by symmetry.}
\end{figure}

The relation between the phase diagram of the out-of-equilibrium transition in the $\Delta-H$ plane and the behavior 
of the quenched complexity $\Sigma_Q(m;H)$ as a function of $m$ is summarized schematically in Fig.~\ref{Fig20} for different values of $H$ (for simplicity, we only consider $H>0$). The solid line represents the coercive field $H_{coer}(\Delta)$ at which the
macroscopic discontinuity in the magnetization occurs. This line ends at the critical point (CP). 
In the strong-disorder regime (points E and F), the complexity has a single maximum that moves towards positive magnetizations as $H$ increases. This behavior is qualitatively similar to the one observed in 1 dimension (e.g., on random graphs with $z=2$). In the low disorder regime, the behavior as $H$ increases is more complicated. 
Whereas $\Sigma_Q(m)$ has a single maximum at $m=0$ when $H=0$ (point A), there are two maxima when $H$ approaches the coercive field (point B). Moreover, the global maximum is the one with the
smallest magnetization, which is negative (although the field is positive). At a certain field, 
close to but smaller than $H_{coer}$, the maximum on the positive side of the magnetizations becomes the global one (which thus corresponds to a discontinuity in the typical magnetization of the states). At $H=H_{coer}$ (point C),  all the states  have a positive magnetization except one. When $H>H_{coer}$ (point D), the complexity has a single maximum at a large positive magnetization.

Note that we have not detailed the passage from one maximum to two maxima in $\Sigma_Q(m)$ as $H$ increases. This point is still unclear and deserves further investigation. For simplicity, we have also assumed that $\Sigma_Q(m)$ goes continuously to zero when $m$ approaches the borders of the gap (i.e., when $m \rightarrow m^{-}_{+\infty}$ or $m\rightarrow m^{+}_{-\infty}$, see Figs.~\ref{Fig1_5} and \ref{Fig20}). However, this is not mandatory. Note also that the behavior for larger connectivities may be more complicated than the one depicted in Fig.~\ref{Fig20}. Indeed, when $z\rightarrow \infty$, one expects to recover the mean-field behavior with its distinct $S$ shape (which in turn implies that $\Sigma_Q(m;H=0)$ must have three maxima when $z$ is large). This is in fact a very interesting issue which justifies an analytical study of the quenched complexity on random graphs in the large-$z$ limit. This study should also permit to understand the actual status of the ``unstable'' branch that can be computed from the fixed-point equation of Ref.~\onlinecite{Dhar1997}. The present numerical study seems to indicate that this branch does not coincide with the boundary $\Sigma_Q(m;H)=0$ but finite-size effects are too important to yield a definite conclusion.

Finally, it would be interesting (but probably highly nontrivial) to study the behavior of the quenched complexity as the critical point CP is approached. It is indeed remarkable that the essential phenomenology of the out-of-equilibrium phase transition (including the critical behavior) is encoded in $\Sigma_Q(m,H$).

\section{Acknowledgements}
We acknowledge fruitful discussions with L. Truskinovsky and E. Vives.  FJPR was
supported by the Marie-Curie contract MRTN-CT-2004-505226 and by the spanish grant MEC EX2005-0792.
The LPTMC is the UMR 7600 of the CNRS.

%\bibliography{../bibliography/bibliography}

\begin{thebibliography}{10}
\bibitem{note1} $\mathbf{s}'\geq \mathbf{s}$ if $s_i'\geq s_i,\; \forall i$.  The ferromagnetic interaction ensures
that this order relation is preserved by the dynamics during a monotonous field history (i.e., if $H$ increases from $H$ to $H'$, then $\mathbf{s}(H') \geq \mathbf{s}(H)$). This is the so-called no-passing rule\cite{Sethna1993}.
\bibitem{note2}The behavior of $\overline{{\cal N}(m;H,N)}$ in the thermodynamic limit is expected to be dominated by the exponential $e^{N\Sigma_A^{\infty}}$.  Accordingly, for finite systems, we assume that $\overline{{\cal N}(m;H,N)}=\exp[N\Sigma_A^{\infty}+a \ln N +b]$, introducing the natural subdominant dependencies associated with an exponential behavior.  The ansatz for $\Sigma_A(m;H,N)$ is readily obtained by using this expression in Eq. (\ref{Complex.Eq.1}).
\bibitem{note3} For instance, one can compute the quenched complexity associated to configurations that are not fully metastable (i.e., with a ``misalignment`` parameter $\mu>0$). Interestingly, we have found that the gap in magnetization also concerns the states with $\mu$ sufficiently small. We plan to come back to this issue in a future work.
\end{thebibliography}

\end{document}